\documentclass[aps,prd,preprint,groupedaddress,nofootinbib]{revtex4-1}

\usepackage{amsmath}
\usepackage{graphicx}
\usepackage{amssymb}
\usepackage{amsfonts}
\usepackage{verbatim}
\usepackage{color}
\usepackage{subfigure}
\usepackage{hyperref}
\usepackage{hhline}
\usepackage{dcolumn}

\def\beq{\begin{eqnarray}}
\def\eeq{\end{eqnarray}}

\def\ga{\gamma}

\begin{document}


\title{Torsion Limits From $t\bar{t}$ Production at the LHC}
\vskip 6mm




\author{F. M. L. de Almeida Jr.}
\email[]{fmarroquim@ufsj.edu.br}
\author{F. R. de Andrade}
\author{M. A. B. do Vale}
\email[]{aline@ufsj.edu.br}
\affiliation{Departamento de Ci\^encias Naturais, Universidade Federal de S\~ao Jo\~ao del-Rei,\\
S\~ao Jo\~ao del Rei, 36301-160, MG, Brazil}

\author{A. A. Nepomuceno}
\email[]{asevedo@gmail.com}
\affiliation{Departamento de Ci\^encias da Natureza, Universidade Federal Fluminense\\
Rua Recife, s/n, Rio das Ostras, RJ, 28890-000, Brazil}



\date{\today}

\begin{abstract}
\noindent
Torsion models constitute a well known class of extended quantum gravity models. 
In this work, one investigates the phenomenological consequences of a torsion field interacting  
with top quarks at the LHC. A torsion field could appear as a new heavy state characterized by 
its mass and couplings to fermions. This new state would form a resonance decaying into a top anti-top pair. 
The latest ATLAS $t\bar t$ production results from LHC 13 TeV data are used to set limits on torsion parameters. 
The integrated luminosity needed to observe torsion resonance at the next LHC upgrades are also evaluated, 
considering different values for the torsion mass and its couplings to Standard Model fermions. 
Finally, prospects for torsion exclusion at the future LHC phases II and III are obtained using
fast detector simulations. 

\end{abstract}


\maketitle

\section{Introduction}

Since the start of the LHC era, its experiments have performed an impressive set of measurements 
in different energy regimes that has strengthened our confidence in the Standard Model (SM).  
The most significant result was the discovery of the Higgs boson by the ATLAS and CMS experiments \cite{Higgs_ATLAS, Higgs_CMS}. 
On the other hand, new physics has not yet appeared, and so far we have no indications from the data on what kind of 
physics beyond SM lies in the high energy scales. Drastic experimental constraints have been put in many extensions of SM, 
making some of them less appealing. As a result, the famous SM puzzles that motivated the pre-LHC model building 
era are still unsolved.  

The various SM extensions deal with one or more problems of the theory, but few of them try to 
incorporate quantum gravity. In fact, it is generally accepted that a consistent quantum gravity theory does not exist. 
In this scenario, the most realistic candidate to a universal theory would be the string theory, 
which induces gravitational interactions in the low energy limit. However, there is no perspective in near future that 
this theory could be verified experimentally. The alternative has been to apply effective approaches to the problem by 
considering natural extension of General Relativity (GR), assuming they might come from a still unknown fundamental theory. 
One of the most natural extensions of GR is the torsion gravity theory \cite{hehl76}. 

There are different approaches to treat the torsion field, but for the purpose of this paper, 
we consider the one where torsion is a fundamental propagating field, with a well-defined action and 
characterized by a mass and couplings between torsion and fermions \cite{shapiro2002}. 
As the torsion is taken as a dynamical field, it is incorporated into the SM along with the other vector fields. 
The coupling between torsion and fermions can be, in principle, non-universal. This possibility is explored in this paper. 

At the LHC, torsion signals could be observed through its decay into fermions. 
The LHC theoretical reach to probe torsion was investigated in \cite{nepomuceno2009,aline2007}. Limits on torsion parameters have 
been derived using LEP and TEVATRON results \cite{belyaev1998, belyaev1999}. The ATLAS experiment 
has put limits on torsion mass and couplings using 7 TeV data in dilepton channel \cite{atlas2012}. 
The impact of a heavy torsion on top pair asymmetries was studied 
in \cite{ayazi2013}.
In most of these studies, the torsion-fermion couplings were considered universal.
Limits on torsion coupling to a scalar field, when it is identified as a dark matter candidate, 
were derived in \cite{belyaev2017}.

In this paper, we use the latest ATLAS $t\bar{t}$ production results to constrain torsion parameters 
assuming representative values for the torsion-top coupling. Limits on torsion from this channel, using LHC 
published data, are derived 
here for the first time. The torsion discovery potential and exclusion at LHC Run II and III are also evaluated. 
Torsion decaying into $t\bar{t}$ at LHC was first investigated in \cite{aline2007}, 
but the subsequent top decay and the final state reconstruction were not taken into account. 
In the present study, we go a step further in understanding the actual collider 
signature by considering measurable final states.

This article is organized as follows. In section \ref{sec:theory} a very brief review of torsion model is given, 
highlighting the features that are most relevant to the current analysis. Section \ref{sec:mc} describes the Monte Carlo 
and detector simulation procedures. In section \ref{sec:limits}, experimental bounds on torsion mass and couplings are derived 
from ATLAS published 13 TeV LHC collision data. 
Section \ref{sec:discovery} presents a fast detector simulation performed to obtain
the LHC potential to observe torsion resonances decaying into $t\bar{t}$. Prospects for torsion exclusion at next LHC 
upgrades are also presented. Conclusions are drawn in Section \ref{sec:conclusions}.

\section{\label{sec:theory}The Torsion Interactions}

The interaction between a Dirac field and torsion, assuming that the metric is flat, is described by the 
following action \cite{belyaev1999}

\beq
{\cal S}_{non-min}^{TS-matter} 
\,=\, i\,\int d^4x\,\,{\bar \psi_{(i)}}\,
\Big(\, \ga^\mu\,\partial_\mu  
+ i\eta_{i}\ga^\mu \ga^5 S_\mu
- im_i \,\Big)\,\psi_{(i)}\,,
\label{tsfermions}
\eeq

\noindent
where $\psi_{(i)}$ stands for each of the SM fermions, $S_{\mu}$ is a axial vector field and $\eta_{i}$ 
is the non-minimal interaction parameter for the corresponding spinor. 
The spinor-torsion interaction enter the SM as interactions of fermions with the new axial vector $S_{\mu}$, 
characterized by new dimensionless parameter, the coupling constants $\eta_i$. 

Unitarity and renormalization conditions in the effective low-energy quantum theory lead to the  torsion action of the form 

\beq
{\cal S}^{\normalfont{TS-Free}} 
\,=\, \,\int d^4x
\Big(-\frac{1}{4}S_{\mu\nu}S^{\mu\nu} + \frac{1}{4}M_{TS}^2S_{\mu}S^{\mu} \Big) .
\label{tsfree}
\eeq

\noindent
where $M_{TS}$ is the torsion mass and $S_{\mu\nu} = \partial_{\mu}S_{\nu} - \partial_{\nu}S_{\mu}$. 
To preserve unitarity, it has been shown that the following relation must be satisfied for each fermion 
of mass $m_i$ \cite{berredo2000}

\beq
\frac{M_{TS}^2}{\eta_i} \gg m_{i}^2
\label{mtsferm}
\eeq

The torsion-fermion interactions are not necessary universal since the values of the couplings $\eta_i$ may not be the same 
for all fermions. The difference comes from the renormalization group equations for each $\eta_i$ that depend on the 
Yukawa coupling for the corresponding fermion. From simplified assumptions \cite{aline2007} it is possible to conclude that, at TeV scale, 
the values of all $\eta_i$ must be the same, except for the top quark. The torsion-top  coupling, denoted hereafter by $\eta_t$, 
may be different because of the potentially stronger running between the Planck and TeV scales.  
Hence, the free parameters of the theory include $M_{TS}$, 
the torsion-top coupling $\eta_t$ and the coupling between torsion and all other SM fermions, denoted by $\eta_f$. 
In order to be as general as possible, using these assumptions, we explore in our analyses the parameter space regions where  
$\eta_t <  \eta_f$,  $\eta_t =  \eta_f$ and $\eta_t >  \eta_f$. 

\section{Monte Carlo and Detector Simulation}
\label{sec:mc}
The torsion effective model was implemented in \textsc{calchep} event generator \cite{belyaevCHEP} according to Eq. (\ref{tsfermions}). 
The implementation was validated and tested for consistency and unitarity.
\textsc{calchep} is used to calculate cross sections and to generate events in which a torsion resonance is produced in 
proton-proton collisions and decays into a pair of top quarks. The simulation is done using the MSTW2008nlo 
parton distribution function \cite{mstw2008} and center-of-mass energy of 13 TeV. The generated events are processed by 
\textsc{pythia 8} \cite{pythia8} for hadronization and decays. 
\textsc{pythia 8} is also used to generate SM $t\bar t$ background events.
For this study, the semileptonic decay of the top quark is selected. 
A fast detector simulation is performed using \textsc{delphes} \cite{delphes} with ATLAS detector configuration,
but pile-up is not taking into account.
The torsion masses were taken in the range 500 GeV to 4500 GeV. 
The coupling $\eta_f$ varied from 0.1 to 0.5 in steps of 0.1. 
Three representative values $\eta_t$ was chosen: 0.1, 0.5 and 1.0. 
These coupling values were selected to produce significant variation 
on $\sigma(pp \rightarrow TS \rightarrow t\bar t)$ without violating 
the constraint imposed by Eq. (\ref{mtsferm}). 

\begin{figure}
 \begin{center}
\includegraphics[width=90mm]{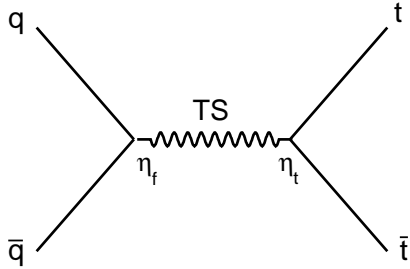}
\caption{\label{fig:feynman} Torsion contribution to $t\bar t$ production at LHC.}
\end{center}
\end{figure}

\begin{figure}
\centering     
\subfigure[]{\label{fig:xsecA}\includegraphics[width=80mm]{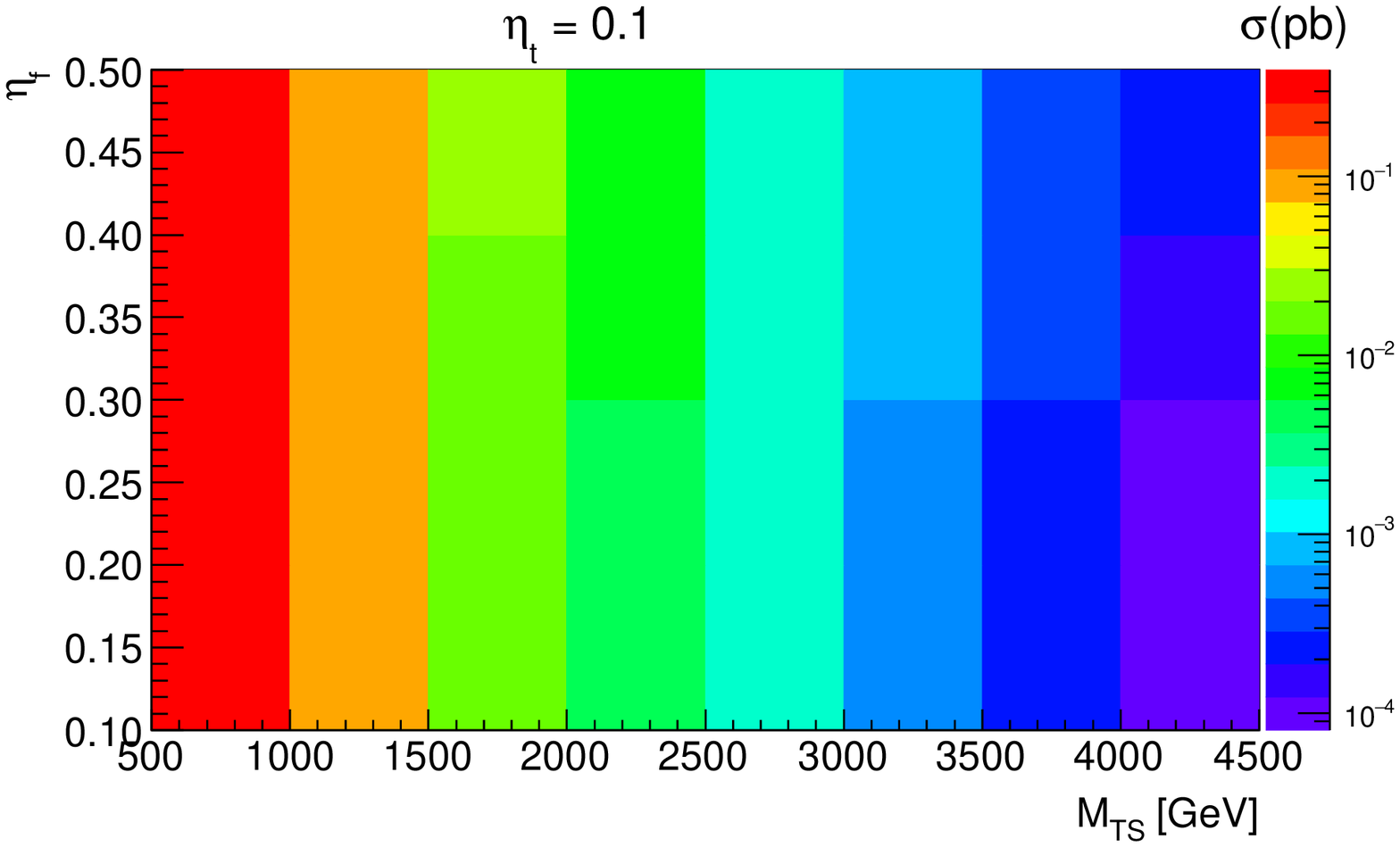}}
\subfigure[]{\label{fig:xsecB}\includegraphics[width=80mm]{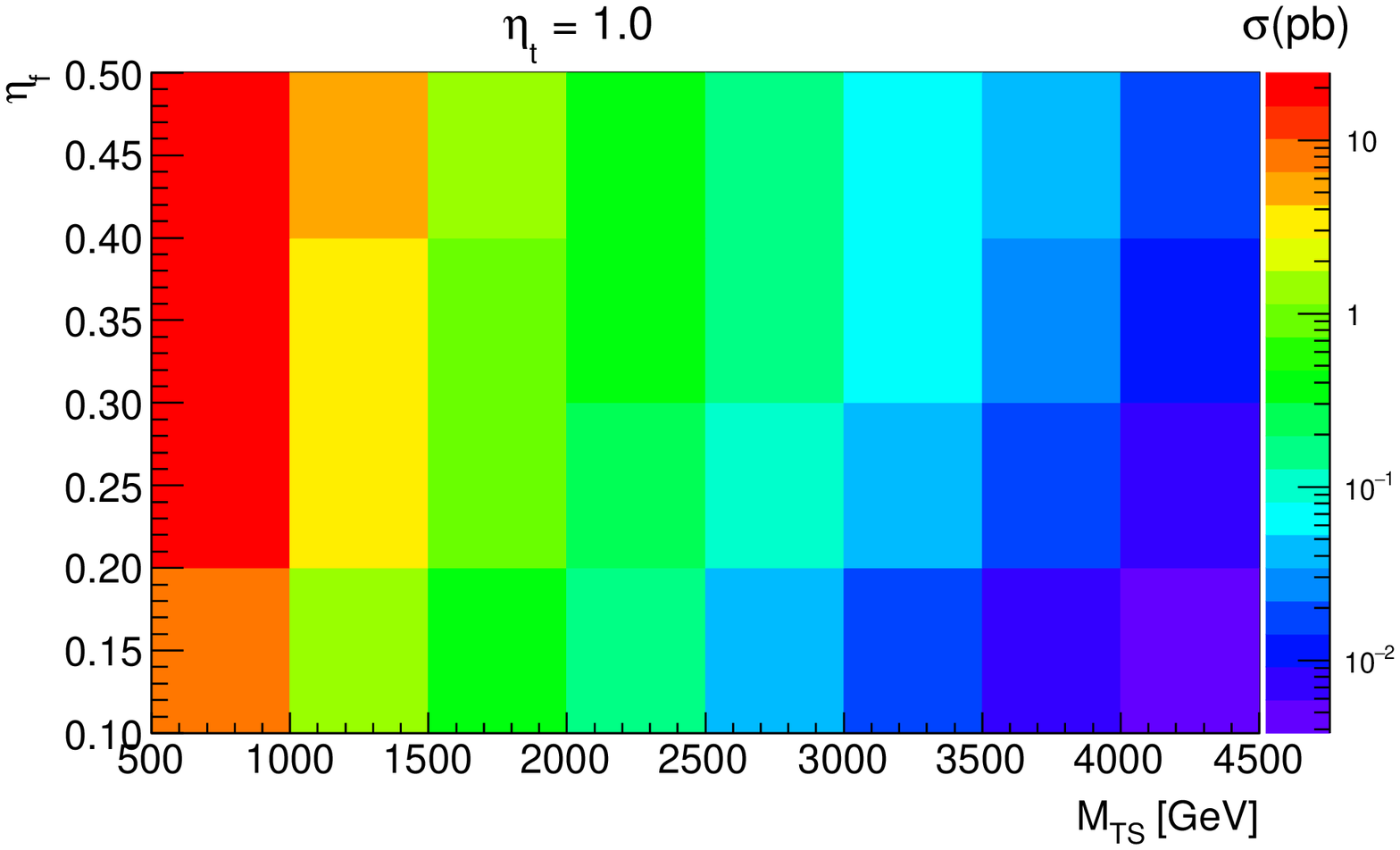}}
\caption{\label{fig:xsec}
LO cross section for process $pp \rightarrow TS \rightarrow t\bar t$ 
as function of $M_{TS}$ and $\eta_f$ for $\eta_t = 0.1$ $\eta_t = 1.0$.}
\end{figure}

A torsion signal can be observed at the LHC as an increase in the number of $t\bar t$ events 
produced via s-channel torsion exchange. The Feynman diagram of the process is shown in Fig. \ref{fig:feynman}.
The LO cross-sections at 13 TeV, as function of torsion mass and $\eta_f$, are shown in 
Figures \ref{fig:xsecA} and \ref{fig:xsecB} for $\eta_t = 0.1$ and $\eta_t = 1$, respectively.
In the analysis that will follow, a K-factor of 1.3 is applied to the LO cross-sections to account for NLO effects. 

As we can see in Fig.\ref{fig:xsecA}, for $\eta_t = 0.1$, 
$\sigma(pp \rightarrow TS \rightarrow t\bar t)$ does not change significantly 
with $\eta_f$. In this case, the torsion production increases with $\eta_f$, 
but it is compensated by decrease of 
$Br (TS \rightarrow t\bar t)$. Changes in the cross-section 
as a function of $\eta_{f}$ are only important when $\eta_t  > \eta_f$, as illustrated in Fig. \ref{fig:xsecB}.
For this reason, different values of $\eta_f$ are considered in the next sections only 
when $\eta_t = 0.5$ and $\eta_t = 1.0$.

\section{Observed Exclusion Limits}
\label{sec:limits}

The ATLAS experiment has searched for heavy particles decaying into $t\bar t$ at center-of-mass energy of 13 TeV 
with a data sample corresponding to an integrated luminosity of 3.2 $\textnormal{fb}^{-1}$ \cite{atlas_ttbar2016}. 
The analysis selected events where the top and the 
anti-top quarks decay to $W$ boson and bottom quarks ($t \rightarrow W^{+}b$, $\bar t \rightarrow W^{-}\bar b$ ), 
one of the $W$'s 
decays to leptons and the other decays to quarks, 
forming the lepton-plus-jets topology. The number of selected events from data and from different SM processes 
estimated by the experiment, in the electron-plus jet channel ($e$ + jets), is listed in Table \ref{tab:table1}. 
As we can see, there is a deficit of events in data compared to total expected background, 
but it is still compatible with the prediction within the uncertainty. 

\begin{table}[b]
\caption{\label{tab:table1}%
Number of events from data and expected background after the $e$ + jets selection obtained by 
ATLAS Experiment \cite{atlas_ttbar2016}.}
\begin{tabular}{ c @{\qquad} c }
\toprule
\textrm{}&
\textrm{Event Yield}\\
\colrule
$t\bar t$   & 3000 $\pm$ 700 \\
$W$ + jets  &  200 $\pm$ 140 \\
$Z$ + jets  &   33 $\pm$  12 \\
Multi-jet   &  130 $\pm$  70 \\
Diboson     &   46 $\pm$  11 \\
TOTAL       & 3700 $\pm$ 800 \\
\botrule
Data        & 3352  \\
\botrule
\end{tabular}
\end{table}

\begin{figure}
\centering     
\subfigure[]{\label{fig:limiteta05}\includegraphics[width=80mm]{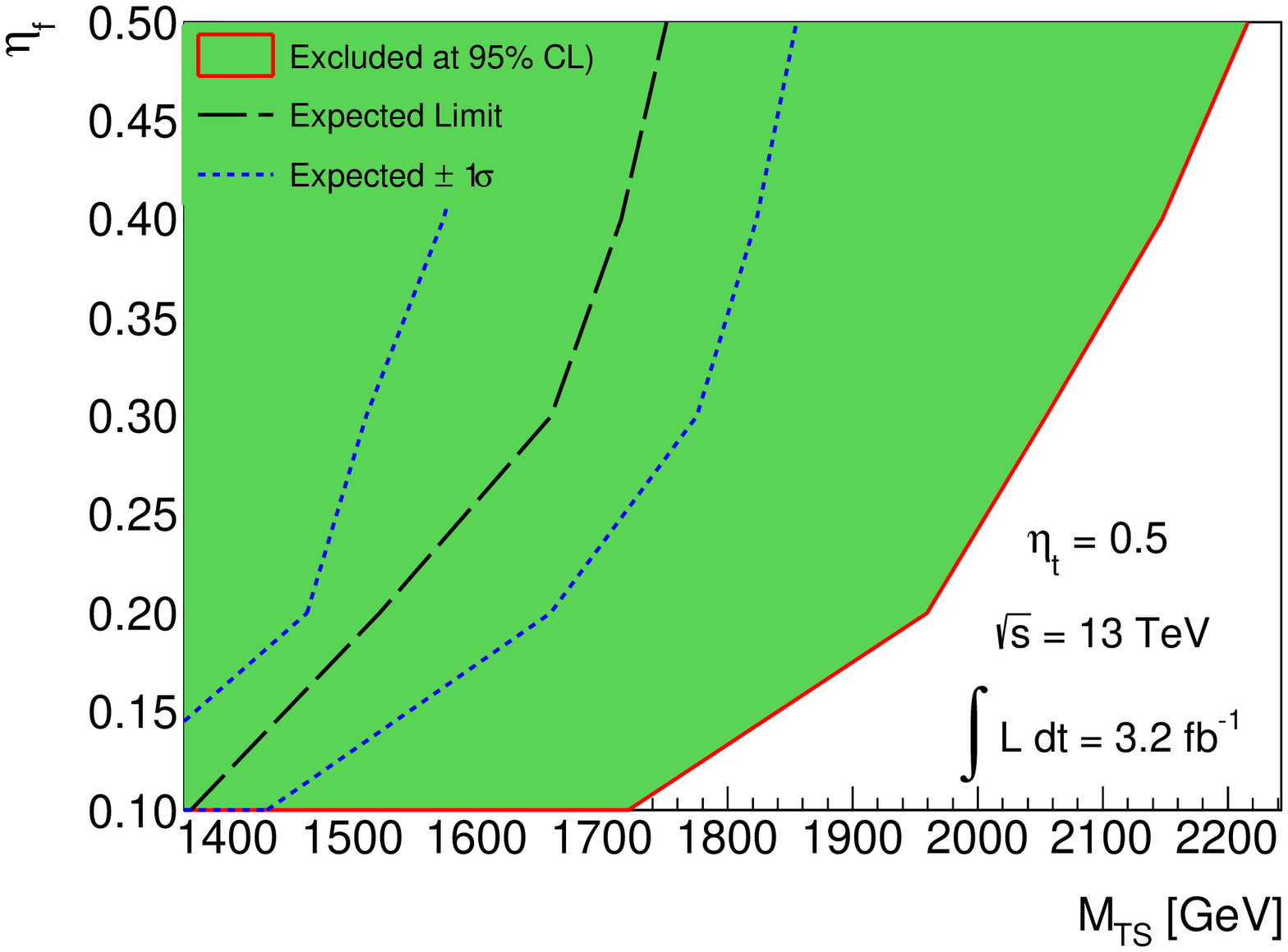}}
\subfigure[]{\label{fig:limiteta1}\includegraphics[width=80mm]{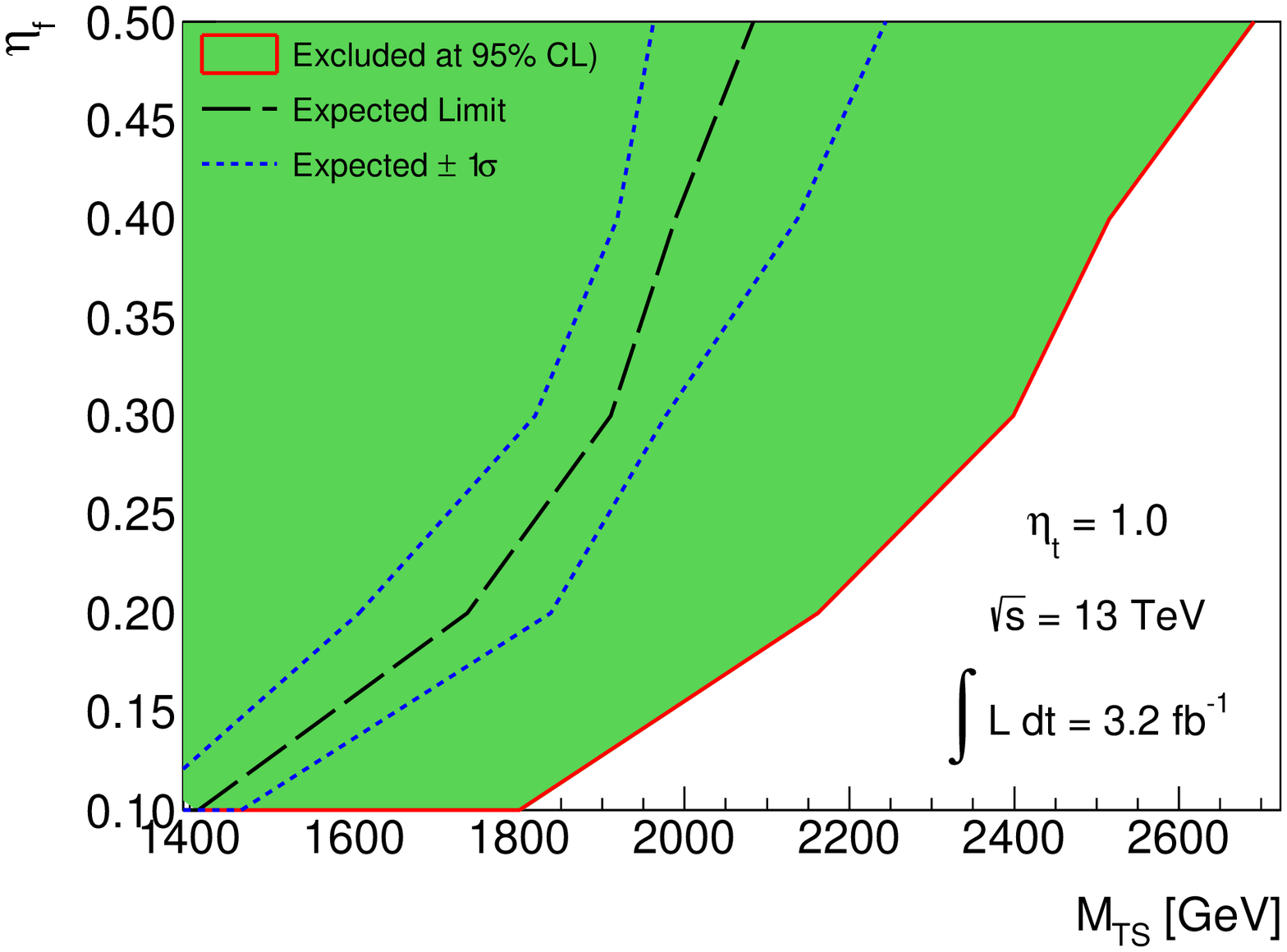}}
\caption{\label{fig:tslimit}
Exclusion region on $M_{TS} \times \eta_{f}$ plane for $\eta_{t} = 0.5$ (a) and 
for $\eta_{t} = 1.0$ (b), based on ATLAS 13 TeV resonance search. The red, long dashed black and dashed blue curves are the observed limits, expected limits and $1\sigma$ error bands, respectively. The green areas are excluded at 95\% C.L. }
\end{figure}

The results from Table \ref{tab:table1} and the theoretical cross-sections calculated in Sec. \ref{sec:mc} are used to set 
limits on torsion mass and couplings. The ATLAS acceptance times efficiency in the $e$+jets channel is also used to 
calculate the number of torsion signal events. Upper limits on the signal cross-section are obtained by applying a Bayesian technique implemented in the 
\textsc{mclimit} program \cite{mclimit, heinrich}. This approach assumes that the signal adds incoherently to the background. 
The inputs for the calculations are the number of events observed in data, the expected number of torsion  
events and the expected number of background events. 
The limit on the cross-section is translated in the lower limit on the torsion mass for each combination of $\eta_{f}$ and $\eta_{t}$. 
Figures \ref{fig:limiteta05} and \ref{fig:limiteta1} show the 95\% C.L. exclusion regions 
on the $M_{TS} \times \eta_{f}$ plane for $\eta_{t} = 0.5$ and $\eta_{t} = 1.0$, respectively.
The large difference between the observed and expected limits is due to the deficit of data events mentioned above. 

In a scenario where torsion is strongly coupled to the top quark but the interaction with other fermions 
is weak ($\eta_{f}$ = 0.1), the current data excludes torsion with a mass between 1700 and 1800 GeV. 
For the highest coupling values considered in this paper, the lower limit on torsion mass is pushed to $\sim$ 2700 GeV. 
For $\eta_{t}$ = 0.1, the observed limit is $M_{TS} <$ 1200 GeV. The observed limits for $\eta_{t}$ = 0.5 and $\eta_{t}$ = 1.0 
are summarized in Table \ref{tab:table2}.

\begin{table}[b]
\caption{\label{tab:table2}%
Observed 95\% C.L. lower limits on torsion mass with varying $\eta_{f}$.}
\begin{tabular}{ c @{\qquad} c  @{\quad} c @{\quad} c @{\quad} c @{\quad} c }
\toprule
\textrm{$\eta_{f}$}&
\textrm{0.1}&
\textrm{0.2}&
\textrm{0.3}&
\textrm{0.4}&
\textrm{0.5}\\
\colrule
Observed Limit [TeV] ($\eta_{t}$ = 0.5)   & 1.72 & 1.96 & 2.05 & 2.15 & 2.22\\
Observed Limit [TeV] ($\eta_{t}$ = 1.0)   & 1.80 & 2.16 & 2.40 & 2.51 & 2.69\\
\botrule
\end{tabular}
\end{table}

\section{Discovery Potential and Limits at Runs II and III}
\label{sec:discovery}
The aim of this section is to perform high mass $t\bar t$ resonance reconstruction and investigate 
the LHC potential to discover torsion at 13 TeV. A fast detector simulation using \textsc{delphes} is performed to determine the 
efficiency for reconstructing the decaying tops from torsion and from SM processes. Jets are reconstructed using the anti-$k_t$ 
algorithm. The missing transverse momentum $E^{miss}_{T}$ is calculated from calorimeter cell energies.

The $t\bar t$ system is reconstructed from the hadronic top ($t \rightarrow Wb \rightarrow b q\bar q$) and from the semi-leptonic top 
($t \rightarrow Wb \rightarrow be\nu_{e}$). The events are required to have exactly one electron with $p_T >$ 30 GeV 
and pseudo-rapidity $\left| \eta \right|<$ 2.5\footnote{$\eta$, without subscription, stands for pseudo-rapidity.}.
Their missing transverse momenta are required to be greater than 30 GeV. The three jets from the hadronic top quark decay can be so collimated 
in the detector that they cannot be distinguished from each other and therefore are reconstructed as a single jet or two jets. 
To select the high boosted top candidates, the following criteria are applied for events with at least two jets: 

\begin{itemize}
	\item the leptonic-top jet is selected by requiring $\Delta R(jet_{\ell}, e) <$ 1.5, where $\Delta R = \sqrt{\Delta \eta^{2} + \Delta \phi^{2}}$ and $jet_{\ell}$ is the jet identified with the expected b-jet from the 
	leptonic top quark. If more than one jet satisfies this condition, the one with highest $p_{T}$ is chosen;
	\item the hadronic-top jet candidate $jet_h$ must be well separated from the lepton by an azimuthal angle distance of $\Delta \phi (jet_h, e) >$ 2.3 rad. Additionally, the monojet $jet_h$ must have $p_T >$ 400 GeV and $\Delta \phi (jet_h, jet_{\ell}) >$ 1.5 rad.
\end{itemize}
 
For the events that do not pass the above selection, three jets are required with one of them having a mass above 70 GeV. It is
assumed that the highest mass jet contains the two merged jets from the $W$ decay or one of $W$ jets merged with the b-jet.
The signal efficiency after these selections depends on $M_{TS}$ and $\eta_f$, and it reaches the maximum value of $\sim$ 2\%. This efficiency already includes the top and $W$ decay branching ratios.
  
\begin{figure}
	\begin{center}
	\includegraphics[width=100mm]{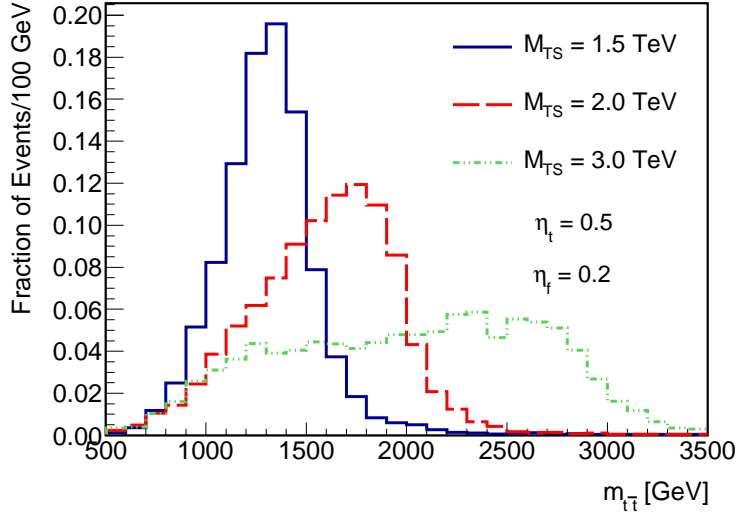}
	\caption{\label{fig:signals} Reconstructed top pair invariant mass distribution for three torsion mass hypotheses. }
	\end{center}
\end{figure}

The invariant mass $m_{t\bar t}$ is reconstructed by adding the four-momentum of the selected semi-leptonic and hadronic top quarks. For the semi-leptonic 
top quark, the longitudinal component of the neutrino momentum is calculated from $E^{miss}_{T}$ and imposing an on-shell $W$ mass constraint. 
When two jets are selected, the monojet $jet_h$ is assumed to be the reconstructed hadronic top quark.
For the events where three
jets are selected, the following $\chi^2$ is used to determine which jet must be assigned to the semi-leptonic or hadronic tops \cite{atlas_ttbar2013,atlas_ttbar2015}:  

\begin{equation}
\chi^2 = \left[ \frac{m_{jj} - m_{t_h}}{\sigma_{t_h}} \right]^2 + \left[ \frac{m_{j\ell\nu} - m_{t_{\ell}}}{\sigma_{t_\ell}} \right]^2
\end{equation}

\noindent
where $m_{t_h}$ $m_{t_\ell}$ are the expected mean masses of hadronic and leptonic top quarks, and $\sigma_{t_h}$ and $\sigma_{t_\ell}$ are their 
respective standard deviations. These four parameters are determined from Monte Carlo simulation studies from \cite{atlas_ttbar2013p}. 
All jet combinations are tested, and the one with the lowest $\chi^2$ value is chosen to calculate $m_{t\bar t}$. Figure \ref{fig:signals} shows the 
reconstructed $m_{t\bar t}$ from the selected events for three torsion mass hypotheses and particular values of torsion-fermions couplings. 
The broad distribution observed for $M_{TS} =$ 3.0 TeV is due to the large resonance width.   

The main background in this analysis is the SM $t\bar t$ production. 
The expected number of background events is estimated using the measured $t\bar t$ cross-section at 13 TeV \cite{atlastop2016} and 
the background event selection efficiency determined from simulation. 
Other backgrounds include production of $W$ and $Z$ bosons associated with jets, single top production, multi-jet 
and di-boson productions. 
We have estimated the number of reconstructed events from these various backgrounds as 20\% of the SM $t\bar t$ reconstructed events. 
Hence, the total number of background events $N_b$ is

\begin{equation}
N_b = 1.2 \times \sigma_{t\bar t}^{SM} \times \epsilon_b \times \mathcal{L}
\end{equation}

\noindent
where $\sigma_{t\bar t}^{SM}$ is the SM $t\bar t$ cross-section, $\epsilon_b$ is the background selection efficiency and $\mathcal{L}$ is the integrated luminosity. 

\begin{figure}
	\centering     
	\subfigure[]{\label{fig:disceta05}\includegraphics[width=80mm]{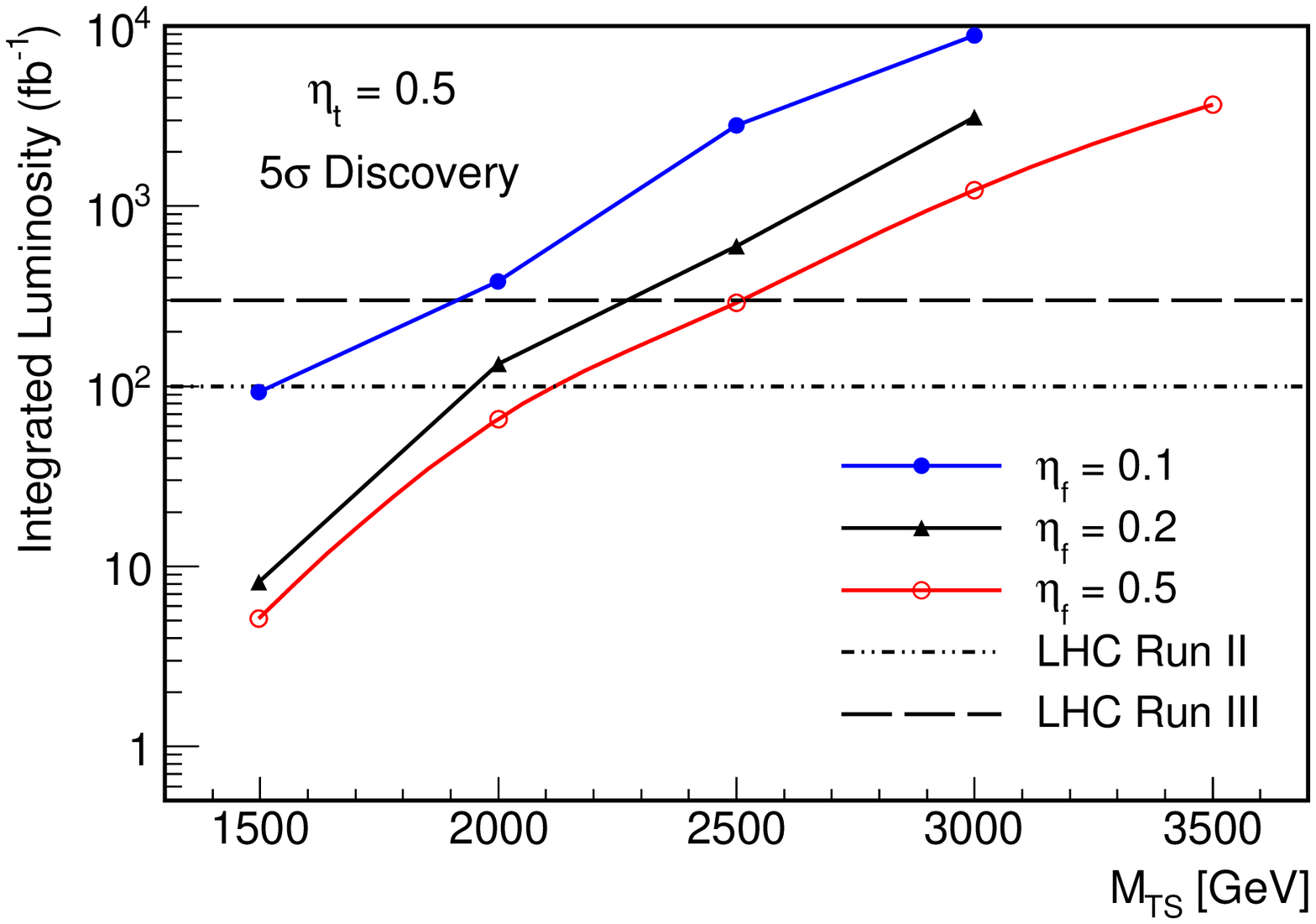}}
	\subfigure[]{\label{fig:disceta1}\includegraphics[width=80mm]{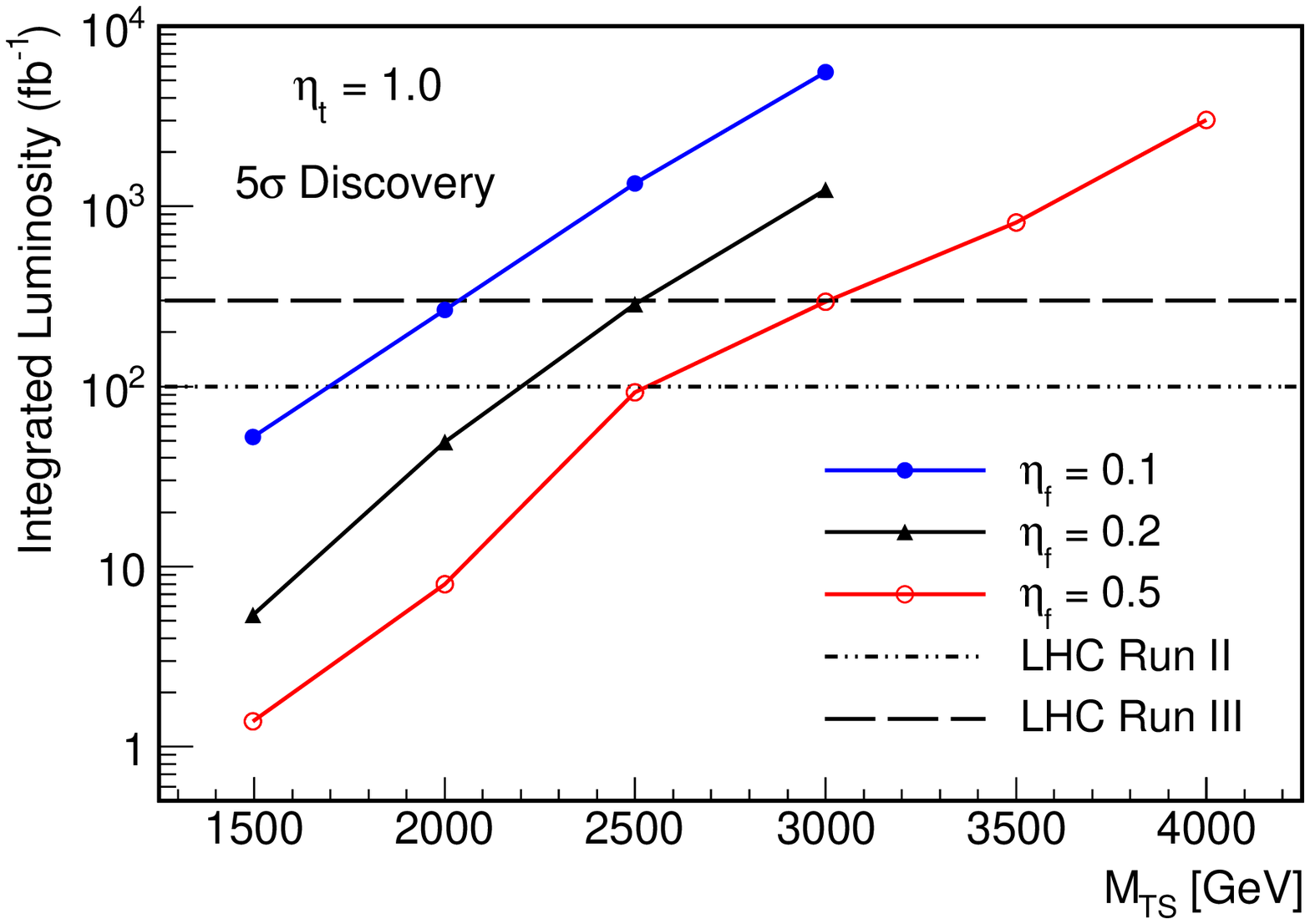}}
	\caption{\label{fig:discovery}
		Experimental sensitivity to observe a torsion signal at LHC 13 TeV. The plots (a) and (b) show the minimal discovering integrated luminosity as a function of $M_{TS}$ for $\eta_t = 0.5$ and $\eta_t = 1.0$, respectively.  }
\end{figure}

In order to determine the LHC experimental sensitivity to probe torsion, the invariant mass $m_{t\bar t}$ 
is used as signal/background discriminant variable. The background is considerably suppressed by applying the cut $m_{t\bar t} > $ 900 GeV.
This selection criteria keeps more than 90\% of the signal events for $M_{TS}  \geq$ 1500 GeV. The background invariant mass distribution, above 900 GeV, is modeled as an exponential using a large simulated MC sample, and a numerical PDF is used to model the signal shape. For a given integrated luminosity $\mathcal{L}$, a likelihood fit is performed to the signal-plus-background  invariant mass distribution in the $m_{t\bar t}$ range of [900,5000] GeV.
The number of signal and background events are the fitted parameters.
Asymptotic formulae for likelihood-based tests \cite{cowan2013} is used to calculate the $P$-value, the probability of a 
background fluctuation being greater than or equal  to the excess observed in the simulated data.
The value of $\mathcal{L}$ is increased and the fitting procedure is repeated until $P$-value $< 3 \times 10^{-7}$, 
the probability required to claim a discovery.  This test is performed for various torsion mass and coupling hypotheses, 
and for each one of them the minimal integrated luminosity needed to discover torsion is obtained. 
The results are shown in Fig. \ref{fig:discovery} for $\eta_t = 0.5$ and $\eta_t = 1.0$.

From Fig. \ref{fig:discovery} we can estimate that by the end of Run II, 
torsion with mass $\sim$ 1500 GeV can be observed at LHC if $\eta_t = 0.5$ and $\eta_f= 0.1$. 
The discovery reach of Run II (Run III) is $\sim$ 2500 GeV ($\sim$ 3000 GeV) if 
torsion is strongly coupled to quarks ($\eta_t = 1.0$, $\eta_f = 0.5$). 
In the high-luminosity LHC scenario ($\mathcal{L}$ = 3000 fb$^{-1}$), 
torsion mass up to $\sim$ 4000 GeV can be probed. For $\eta_t = 0.1$, the maximum discovery reach at LHC is 
$M_{TS} \sim$ 1700 GeV.

\begin{figure}
	\begin{center}
	\includegraphics[width=90mm]{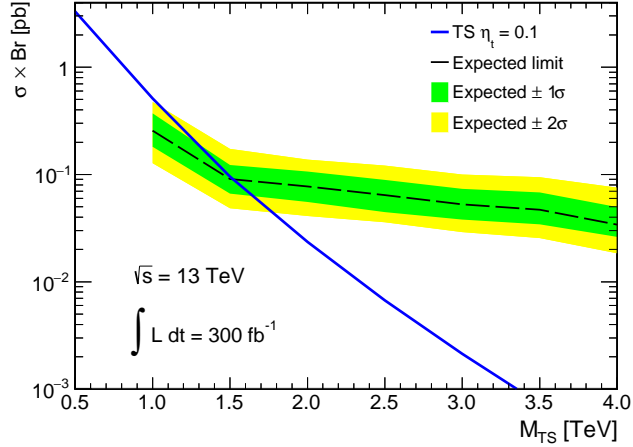}
	\caption{\label{fig:explimitetat01} Upper limit on $\sigma \times Br(TS \longrightarrow t\bar t)$ for $\eta_t$ = 0.1 assuming 
	an integrated luminosity of 300 fb$^{-1}$ for 	diffent torsion masses. The lower limit on torsion mass is 
	obtained from the crossing between the expected limit and the torsion theoretical cross-section. }
	\end{center}
\end{figure}

The invariant mass distribution is also used to calculate torsion expected limits at Runs II and III. 
Expected upper limits on $\sigma \times Br(TS \longrightarrow t\bar t)$ are calculated using simulated pseudo-experiments assuming that only 
SM processes are present. 
A Bayesian approach is used \cite{bat2009}, with a flat prior probability distribution for 
$\sigma Br$. The most probable number of signal events, 
and the corresponding confidence intervals, are determined from a likelihood function defined as the product of the 
Poisson probabilities over all $m_{t\bar t}$ mass bins in the search region, using the appropriate signal invariant mass distribution. 
The limit on the number of events is converted into a limit 
on $\sigma Br$. 95 \% CL upper limit on $\sigma Br$ for each pseudo-experiment is obtained, 
and the median value is chosen to represent the expected limit. This calculation is performed for each combination of the 
parameters ($M_{TS}$, $\eta_f$, $\eta_t$). The 68\% and 95\% error bands are also calculated. 
By comparing the limits on $\sigma Br$ with the torsion theoretical cross-sections, the lower limits on torsion 
mass are determined. 
The procedure result is illustrated in Fig. \ref{fig:explimitetat01} for $\eta_t$ = 0.1 and $\mathcal{L}$ = 300 fb$^{-1}$. 
In this case, the lower bound $M_{TS} <$ 1.52 TeV can be achieved. 

\begin{figure}
	\centering     
	\subfigure[]{\label{fig:limitetat05l100}\includegraphics[width=80mm]{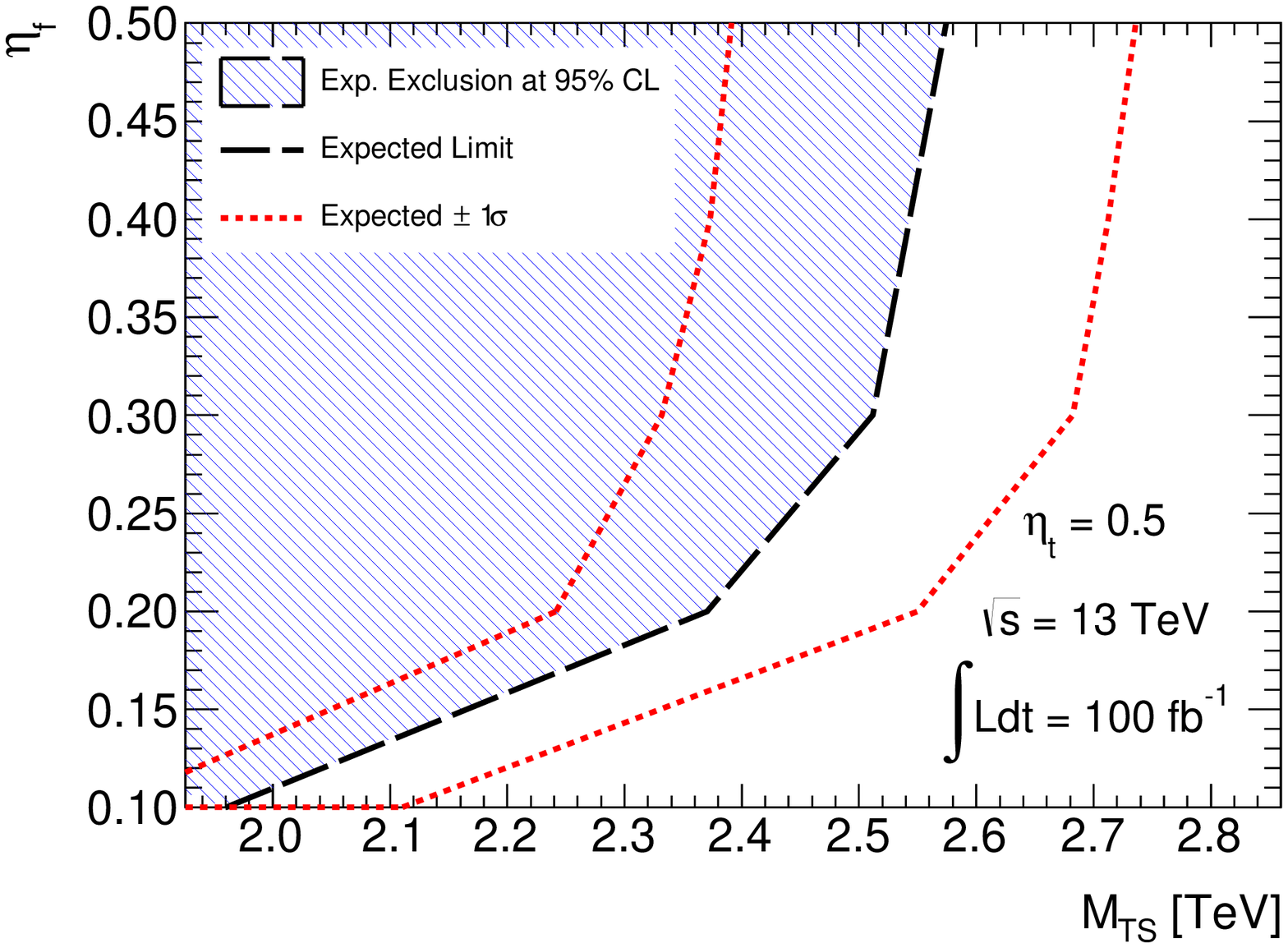}}
	\subfigure[]{\label{fig:limitetat1l100}\includegraphics[width=80mm]{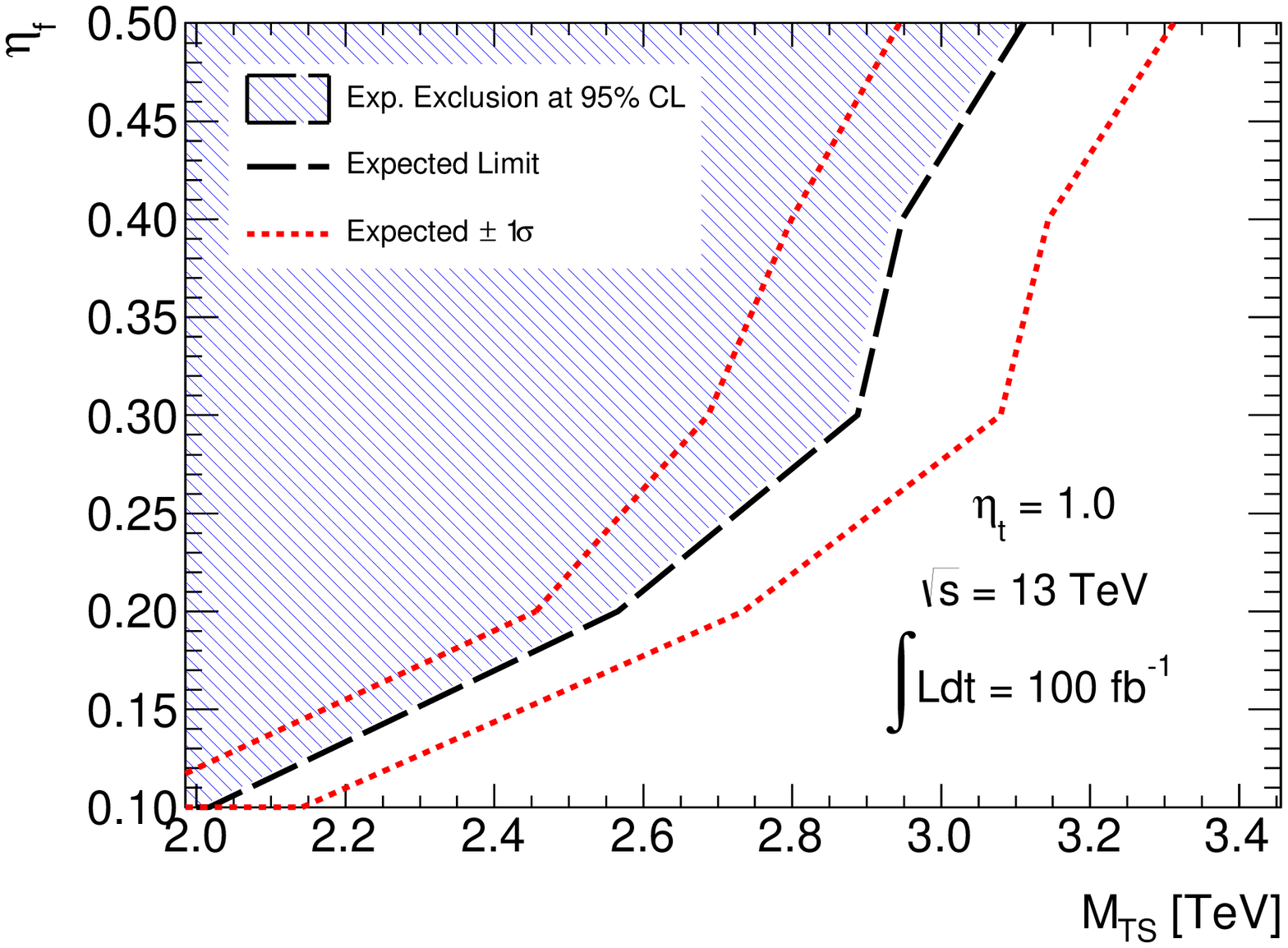}}
	\caption{\label{fig:explimitl100}
              Expected limits on torsion parameters for $\eta_t = 0.5$ (a) and $\eta_t = 1.0$ (b) assuming $\mathcal{L}$ = 100 fb$^{-1}$.
              The long dashed black lines are the expected limits, and the dashed red lines are the 1$\sigma$ variations. 
              The shaded areas would be excluded at 95\% CL. }
		
\end{figure}

\begin{figure}
	\centering     
	\subfigure[]{\label{fig:limitetat05l300}\includegraphics[width=80mm]{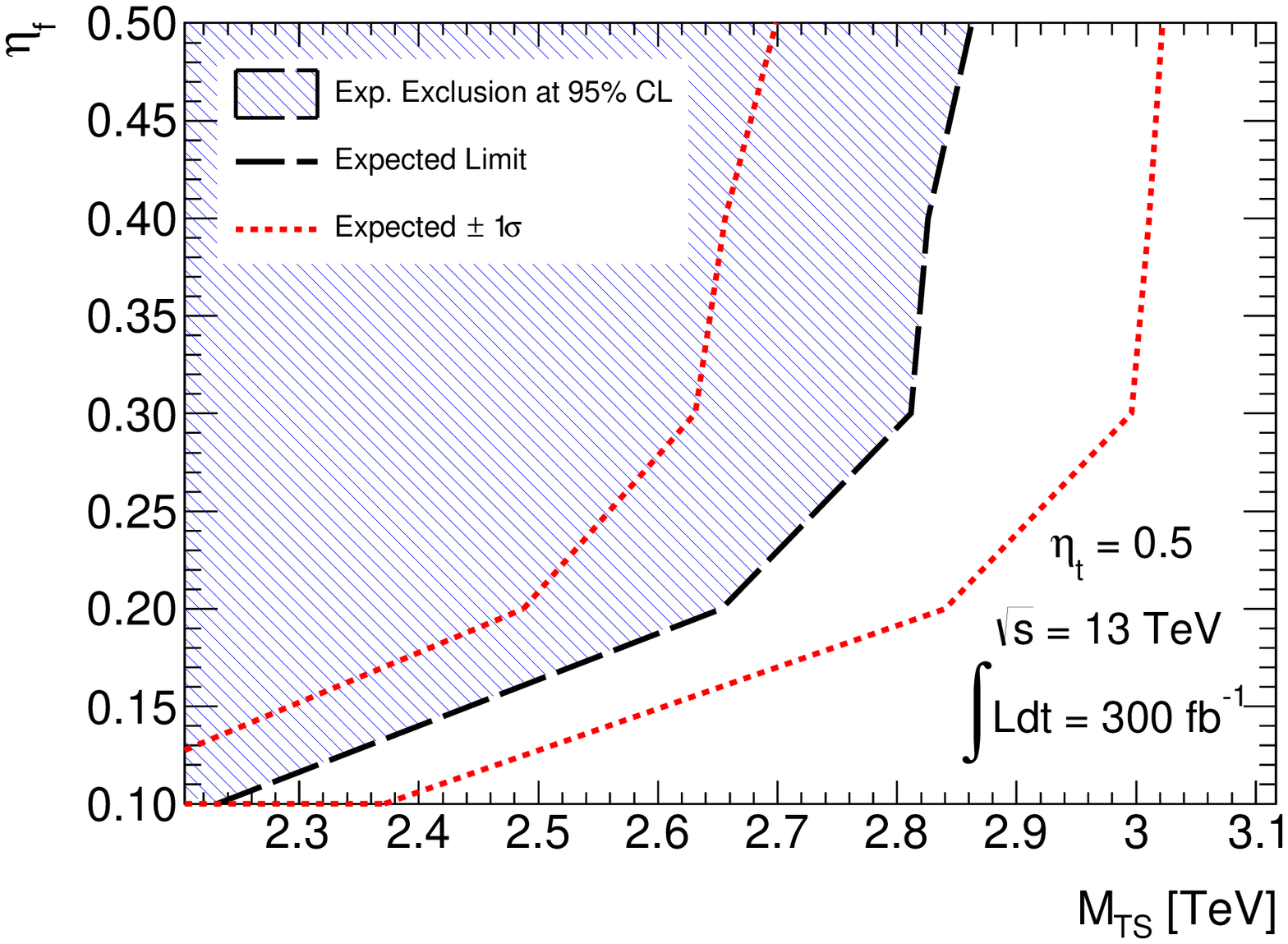}}
	\subfigure[]{\label{fig:limitetat1l300}\includegraphics[width=80mm]{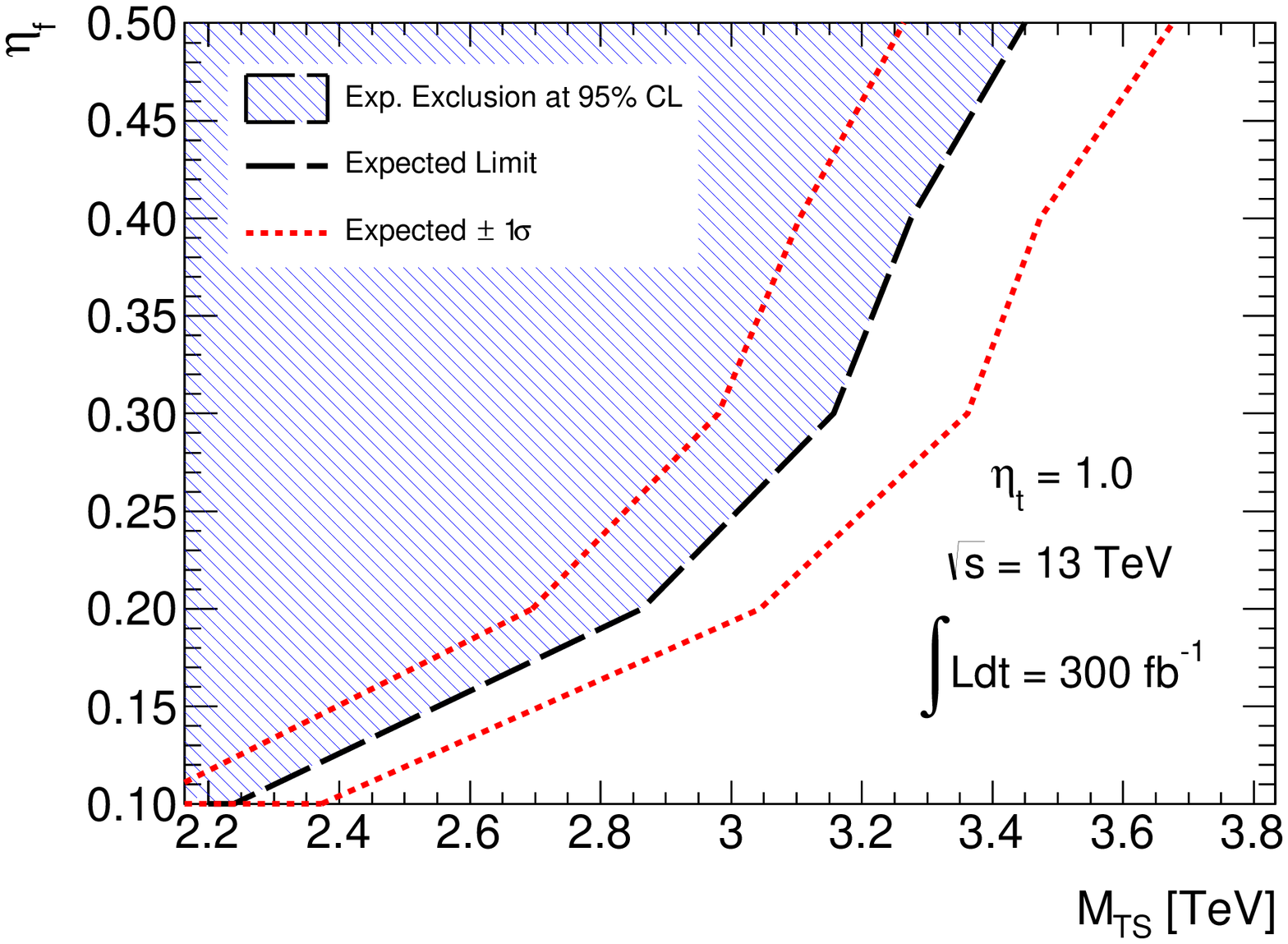}}
	\caption{\label{fig:explimitl300}
           Same as Fig. \ref{fig:explimitl100}, but assuming $\mathcal{L}$ = 300 fb$^{-1}$ }
\end{figure}

Figures \ref{fig:explimitl100} and \ref{fig:explimitl300} show the expected exclusion region on $M_{TS} \times \eta_f$ plane for $\eta_t = 0.5$ 
and $\eta_t = 1.0$ considering $\mathcal{L}$ = 100 fb$^{-1}$ and $\mathcal{L}$ = 300 fb$^{-1}$, respectively. With 100 fb$^{-1}$ of data, 
the torsion mass limits range from 1.96 TeV 
to 3.10 TeV. With $\mathcal{L}$ = 300 fb$^{-1}$, the limits can be extend to $\sim$ 3.5 TeV. 

In Table \ref{tab:table3} we summarize the results obtained for $\eta_t$ = 1.0 for the different integrated luminosities considered. 
The second table column shows the limits obtained in the di-lepton channel by ATLAS Experiment at 7 TeV \cite{atlas2012}. Since di-lepton is a cleaner signature
with better efficiencies, the torsion limits obtained from $t\bar t$ with the current data are still less restrictive. The $t\bar t$ channel, 
however, provides a unique way to test the torsion-top coupling and the torsion non-universality interactions.  
 
\begin{table}[b]
\caption{\label{tab:table3}%
Comparison between the lower limits on $M_{TS}$ obtained by ATLAS experiment at 7 TeV in di-lepton channel 
and the results from $t\bar t$ derived in this paper, for $\eta_t$ = 1.0. 
}
\begin{ruledtabular}
\begin{tabular}{ccccc}
\multicolumn{5}{c}{Lower Limits on $M_{TS}$ at 95\% CL [TeV]}\\
\cline{2-5}\\
\textrm{$\eta_f$}&
\textrm{\shortstack{ATLAS dilepton 7 TeV \\ $\mathcal{L}$ = 5.0 fb$^{-1}$ }}& 
\textrm{\shortstack{$t\bar t$ 13 TeV \\ $\mathcal{L}$ = 3.2 fb$^{-1}$}}&
\textrm{\shortstack{$t\bar t$ 13 TeV \\ $\mathcal{L}$ = 100 fb$^{-1}$}}&
\textrm{\shortstack{$t\bar t$ 13 TeV \\ $\mathcal{L}$ = 300 fb$^{-1}$}}\\
\colrule
0.1 & 1.94 & 1.80 & 2.02 & 2.24\\
0.2 & 2.29 & 2.16 & 2.57 & 2.86\\
0.3 & 2.50 & 2.40 & 2.89 & 3.16\\
0.4 & 2.69 & 2.51 & 2.95 & 3.27\\
0.5 & 2.91 & 2.69 & 3.11 & 3.45\\
\end{tabular}
\end{ruledtabular}
\end{table}

\section{Conclusions}
\label{sec:conclusions}

Exclusion limits on torsion mass and couplings based on ATLAS $t\bar t$ results from LHC 13 TeV data are derived. 
Considering non-universal torsion-quarks couplings, torsion with masses between 1.2 TeV to 2.7 TeV 
are excluded at 95\% CL. The LHC potential to observe torsion decaying 
into $t \bar t$ is also investigated. Taking into account the reconstruction and selection efficiencies of 
the decaying top quarks, 
it is found that torsion with mass up to $\sim$ 3.0 TeV can be observed by the end of Run III. 
The results also show that a torsion mass of 4.0 TeV 
set the maximum $M_{TS}$ value that can be probed at LHC from $t \bar t$ production in the electron-plus-jets selection. 
New data from Run II and Run III can extend the current 
torsion mass limits to $\sim$ 3.5 TeV.  
The results presented in this paper are complementary to torsion searches in di-lepton channel 
and provide information on torsion-top coupling. 

\begin{acknowledgments}
\noindent
This work is supported by the National Council of Research (CNPq) under grants 308494/2015-6 and 402846/2016-8,
and by Funda\c{c}\~ao de Amparo a Pesquisa do Estado de Minas Gerais (FAPEMIG), APQ 03179-15
\end{acknowledgments}

\bibliography{bibliography}
\end{document}